\documentclass[aps,prc,twocolumn,amsfonts,showpacs,%
groupedaddress,floatfix]{revtex4}
\usepackage{bm,amsmath,epsfig,dcolumn,slashbox}

\begin{document}

\title{ An Alternative Parameterization of ${\bm R}$-matrix Theory }

\author{ C.~R.~Brune }

\affiliation{Edwards Accelerator Laboratory,
  Department of Physics and Astronomy,
  Ohio University, Athens OH 45701, USA }

\date{\today}

\begin{abstract}

An alternative parameterization of ${\bm R}$-matrix theory is presented
which is mathematically equivalent to the standard
approach, but possesses features which simplify the fitting of
experimental data. In particular there are no level shifts and no
boundary-condition constants which allows the positions and partial
widths of an arbitrary number of 
levels to be easily fixed in an analysis.
These alternative parameters can be converted to standard ${\bm R}$-matrix
parameters by a straightforward matrix diagonalization procedure.
In addition it is possible to express the collision matrix
directly in terms of the alternative parameters.

\end{abstract}

\maketitle

\section{Introduction}

The ${\bm R}$-matrix theory of reactions  has proven over
the course of time to be very useful in nuclear and
atomic physics, both for the fitting of experimental data
and as a tool for theoretical calculations.
In this paper we explore a mathematically-equivalent
alternative formulation of ${\bm R}$-matrix theory
which will be especially useful for the
fitting of experimental nuclear physics data.

In a recent paper an alternative parameterization 
${\bm R}$-matrix theory was described by Angulo and Descouvemont~\cite{Ang00}.
In their framework there are no level shifts and it is
straightforward to incorporate known information about
level energies and partial widths.
They presented an approximate iterative relation between
the alternative parameters and the standard ${\bm R}$-matrix parameters.
In addition consideration was limited to
the single-channel case with a boundary condition constant of zero.
Some aspects of these alternative parameters
have also been discussed in a paper by Barker~\cite{Bar71}.
In this paper we further develop the concept of
an alternative ${\bm R}$-matrix parameterization.
The description is generalized to allow non-zero boundary condition
constants and an arbitrary number of channels.
We present an exact method for converting the alternative parameters
to the standard ${\bm R}$-matrix parameters which only
requires a matrix diagonalization.
We also found a rather surprising result, that the collision matrix
can be calculated directly from the alternative parameters
using alternative formulations of the level matrix or ${\bm R}$ matrix.
We then discuss the solution of the nonlinear eigenvalue equation
required to extract the alternative parameters from the
standard parameterization, and demonstrate some of these ideas
using a simple example.
Finally we briefly discuss the application of the alternative
parameterization to $\gamma$ rays and $\beta$ decays.

\section{Review of standard ${\bm R}$-Matrix theory}
\label{sec:standardR}

We begin by reviewing some of the notation and results of standard
${\bm R}$-matrix theory as described by Lane and Thomas (LT)~\cite{Lan58}.
The ${\bm R}$ matrix is a function of the energy $E$ and is defined by
\begin{equation}
R_{c'c}=\sum_\lambda \frac{\gamma_{\lambda c'}\gamma_{\lambda c}}
  {E_\lambda -E},
\label{eq:rmatrix}
\end{equation}
where $E_\lambda$ are the level energies,
$\gamma_{\lambda c}$ are the reduced width amplitudes,
$\lambda$ is the level label, and $c$ is the channel label.
We will assume that the numbers of levels and  channels
are finite and given by $N_\lambda$ and $N_c$, respectively.
One must also specify the constants $B_c$,
which determine the boundary conditions satisfied by the
underlying eigenfunctions.

In order to calculate physical observables one must employ
various combinations of the Coulomb wavefunctions,
evaluated at the channel radius $r_c=a_c$.
The quantities $I_c$ and $O_c$ are defined by (LT, Eq.~II.2.13).
For closed channels the outgoing solution $O_c$ is taken to be
the exponentially-decaying Whittaker function (LT, Eq.~II.2.17).
In addition one defines $\Omega_c=(I_c/O_c)^{1/2}$ and
\begin{equation}
L_c=\left(\frac{a_c}{O_c}\frac{\partial O_c}
{\partial r_c} \right)_{a_c} =S_c+{\rm i}P_c
\end{equation}
where the shift factor $S_c$ and penetration factor $P_c$
are real quantities.
The collision matrix ${\bm U}$ is an $N_c\times N_c$ matrix which
determines the observable quantities; it is related to
the ${\bm R}$ matrix via (LT, Eq.~VII.1):
\begin{equation}
{\bm U}=2{\rm i}{\bm \rho}^{1/2}{\bm O}^{-1}[{\bm 1}-{\bm R}
({\bm L}-{\bm B})]^{-1} {\bm R}{\bm \rho}^{1/2}{\bm O}^{-1} +
{\bm I}{\bm O}^{-1}
\label{eq:u_channel}
\end{equation}
where ${\bm O}$, ${\bm I}$, ${\bm L}$, ${\bm B}$, and ${\bm \rho}$
are purely diagonal with elements $O_c$, $I_c$, $L_c$, $B_c$, and
$k_c a_c$, respectively; ${\bm 1}$ is the unit matrix,
and $k_c$ is the wavenumber.

It is convenient to form the level-space column vector
${\bm \gamma}_c$ from the $\gamma_{\lambda c}$, and to then
form the rectangular matrix ${\bm \gamma}$ from the ${\bm \gamma}_c$
such that the matrix ${\bm \gamma}$ has $N_\lambda$ rows and $N_c$ columns.
In addition, the diagonal matrix ${\bm e}$ is defined by
\begin{equation}
e_{\lambda\mu}=E_\lambda \delta_{\lambda\mu}.
\label{eq:little_e}
\end{equation}
The ${\bm R}$ matrix defined by Eq.~(\ref{eq:rmatrix})
can now be written succinctly as
\begin{equation}
{\bm R}={\bm \gamma}^T({\bm e}-E{\bm 1})^{-1}{\bm \gamma}.
\end{equation}
The collision matrix can also be expressed as
\begin{equation}
{\bm U}=2{\rm i}{\bm \rho}^{1/2}{\bm O}^{-1}{\bm \gamma}^T{\bm A}{\bm \gamma}
{\bm \rho}^{1/2}{\bm O}^{-1} + {\bm I}{\bm O}^{-1},
\label{eq:u_a_mat}
\end{equation}
where ${\bm A}$ is an $N_\lambda\times N_\lambda$
matrix defined by its inverse:
\begin{equation}
{\bm A}^{-1}={\bm e}-E{\bm 1}-{\bm \gamma}({\bm L}-{\bm B}){\bm \gamma}^T.
\label{eq:ainv}
\end{equation}
The equivalence of these two forms for the collision matrix is discussed
in (LT, Sec.~IX.1) and in the Appendix.
In addition the elements of the collision matrix connecting
open channels in Eq.~(\ref{eq:u_a_mat}) can also be expressed as
\begin{equation}
U_{c'c}=\Omega_{c'}\Omega_c\left[ \delta_{c'c}+2{\rm i}(P_{c'} P_c)^{1/2}
{\bm \gamma}_{c'}^T {\bm A} {\bm \gamma}_c \right] ,
\label{eq:u_a}
\end{equation}
using the definitions of the Coulomb functions.

An interesting feature of ${\bm R}$-matrix theory is that the collision matrix
is invariant under changes in the $B_c$, provided that the
$E_\lambda$ and $\gamma_{\lambda c}$ are suitably adjusted.
This result remains true even for the case of finite $N_\lambda$~\cite{Bar72}.
The transformation is most easily described using matrix equations
in level space. Let us consider the transformation $B_c\rightarrow B_c'$,
$E_\lambda\rightarrow E_\lambda'$, and
$\gamma_{\lambda c}\rightarrow\gamma_{\lambda c}'$.
One first constructs the real and symmetric matrix ${\bm C}$ defined by
\begin{equation}
{\bm C}={\bm e}-\sum_c{\bm \gamma}_c{\bm \gamma}_c^T(B_c'-B_c),
\label{eq:cmatrix}
\end{equation}
which is diagonalized by the orthogonal matrix ${\bm K}$
such that ${\bm D}={\bm K}{\bm C}{\bm K}^T$,
with $D_{\lambda\mu}=D_\lambda\delta_{\lambda\mu}$.
The necessary transformation of the ${\bm R}$-matrix parameters
is then given by~\cite{Bar72}
\begin{equation}
E_\lambda'=D_\lambda
\label{eq:etrans}
\end{equation}
and
\begin{equation}
{\bm \gamma}_c'={\bm K}{\bm \gamma}_c.
\label{eq:gtrans}
\end{equation}
It is straightforward to verify by substitution into
Eqs.~(\ref{eq:u_a_mat},\ref{eq:ainv}) that these transformations
leave ${\bm U}$ invariant.

\section{The Alternative Parameterization}

\subsection{Definition of the parameterization}
\label{subsec:def_alt}

We begin by defining the real and symmetric matrix ${\bm {\mathcal E}}$:
\begin{equation}
{\bm {\mathcal E}}={\bm e}-\sum_c{\bm \gamma}_c{\bm \gamma}_c^T(S_c-B_c),
\label{eq:ematrix}
\end{equation}
and consider the eigenvalue equation
\begin{equation}
{\bm {\mathcal E}}{\bm a}_i = \tilde{E}_i{\bm a}_i
\label{eq:eigen}
\end{equation}
where $\tilde{E}_i$ is the eigenvalue and
${\bm a}_i$ is the corresponding eigenvector.
Note that ${\bm {\mathcal E}}$
is implicitly dependent upon $\tilde{E}_i$ through $S_c$, so
the eigenvalue problem is nonlinear.
We will assume for convenience that the eigenvectors are normalized
so that ${\bm a}_i^T{\bm a}_i=1$.

Before proceeding further we would like to point out two
important properties of this eigenvalue equation:
(1) The eigenvalues $\tilde{E}_i$ are invariant if the $B_c$ are changed
and the $E_\lambda$ and $\gamma_{\lambda c}$ are changed according to
Eqs.~(\ref{eq:etrans},\ref{eq:gtrans}). This result
is easily shown by substituting Eqs.~(\ref{eq:cmatrix}-\ref{eq:gtrans})
into Eqs.~(\ref{eq:ematrix},\ref{eq:eigen}).
(2) If $B_c=S_c(E_\lambda)$, the matrix ${\bm {\mathcal E}}$ is diagonal
for the energy $E_\lambda$ and hence $E_\lambda$ is an eigenvalue.
For this choice of $B_c$ the ${\bm R}$-matrix level energy $E_\lambda$ is
often taken to be the ``observed resonance energy''.
This definition is particularly useful in the present context and
we will thus adopt the $\tilde{E}_i$ as the observed resonance energies.
The $\tilde{E}_i$ also correspond exactly to the level energies
found using boundary-condition constant transformations
yielding $B_c=S_c(E_\lambda)$ such as described
by Barker~\cite{Bar71} and Azuma {\em et al.}~\cite{Azu94}.

In addition one can define a new set of reduced width parameters
$\tilde{\gamma}_{ic}$ via
\begin{equation}
\tilde{\gamma}_{ic}={\bm a}_i^T {\bm \gamma}_{c}.
\label{eq:newgamma}
\end{equation}
These new reduced width parameters are also invariant under
changes in $B_c$. When  $B_c=S_c(E_\lambda)$ we have also
$\tilde{\gamma}_{\lambda c}=\gamma_{\lambda c}$.
The quantities $\tilde{E}_i$ and $\tilde{\gamma}_{ic}$ can be
taken as an alternative parameterization of ${\bm R}$-matrix theory.
We will derive below efficient methods to convert $\tilde{E}_i$ and
$\tilde{\gamma}_{ic}$ into the standard ${\bm R}$-matrix parameters
$E_\lambda$ and $\gamma_{\lambda c}$, or to the collision matrix ${\bm U}$.
Also note that $\tilde{E}_i$ and $\tilde{\gamma}_{ic}$ are equivalent
to the ``superscript $(\lambda)$'' parameters of Barker~\cite{Bar71},
and essentially equivalent to the ``observed'' ${\bm R}$-matrix parameters
described by Angulo and Descouvemont~\cite{Ang00}.

Our Eq.~(\ref{eq:eigen}) is closely related to the complex eigenvalue
equation introduced by Hale, Brown, and Jarmie~\cite{Hal87} to locate
the poles of the collision matrix -- in fact it is just the real part of their
eigenvalue equation.
For bound states our $\tilde{E}_i$ are thus equivalent to
the eigenvalues discussed in Ref.~\cite{Hal87} since $P_c=0$.
For these states we can also introduce the asymptotic normalization
constant $C_{ic}$ which is given by~\cite{Muk99}
\begin{equation}
C_{ic}^2=\frac{2\mu_c a_c}{\hbar^2O_c^2}\left[ \frac{\tilde{\gamma}_{ic}^2}
{1+\sum_c\tilde{\gamma}_{ic}^2\left(
\frac{dS_c}{dE}\right)_{\tilde{E}_i}}\right],
\end{equation}
where $\mu_c$ is the reduced mass. This quantity is simply related
to the pole residues described by Eq.~(4) of Ref.~\cite{Hal87}.
For unbound states there appears to be no simple relation between
$\tilde{E}_\lambda$ and $\tilde{\gamma}_{\lambda c}$ and
the pole parameters of Ref.~\cite{Hal87}.
One may however define the observed partial width of a level in terms of our
parameters by
\begin{equation}
\Gamma_{i c}=\frac{2P_c\tilde{\gamma}_{ic}^2}{1+\sum_c\tilde{\gamma}_{ic}^2
\left(\frac{dS_c}{dE}\right)_{\tilde{E}_i}}
\label{eq:Gamma}
\end{equation}
see (LT, Eqs.~XII.3.5 and~XII.3.6).
One should bear in mind however that there are many different
definitions of observed resonance energies and widths in use;
generally the differences between definitions are
significant only for broad states.

\subsection{Relation to standard parameters}
\label{subsec:rel_stand}

We will next show the method to convert $\tilde{E}_\lambda$ and
$\tilde{\gamma}_{\lambda c}$ to standard ${\bm R}$-matrix parameters.
It is assumed that the eigenvalues are distinct, so that
$\tilde{E}_i \neq \tilde{E}_j$ provided $i\neq j$.
Note that if this were not the case the levels with the same
$\tilde{E}_i$ could be combined into a single level.
The eigenvectors of Eq.~(\ref{eq:eigen}) are not orthogonal;
using the eigenvalue equation with two different eigenvalues one finds
\begin{equation}
{\bm a}_j^T({\bm{\mathcal E}}_j-{\bm{\mathcal E}}_i){\bm a}_i=
(\tilde{E}_j-\tilde{E}_i){\bm a}_j^T{\bm a}_i,
\end{equation}
where ${\bm{\mathcal E}}_i$ is used to denote the matrix
${\bm{\mathcal E}}$ evaluated for the energy $\tilde{E}_i$.
Using Eqs.~(\ref{eq:ematrix},\ref{eq:newgamma})
with this result we obtain
\begin{equation}
{\bm a}_j^T{\bm a}_i=-\sum_c\tilde{\gamma}_{ic}\tilde{\gamma}_{jc}
\frac{S_{ic}-S_{jc}}{\tilde{E}_i-\tilde{E}_j},
\end{equation}
where $S_{ic}$ denotes the shift function $S_c$ evaluated at $\tilde{E}_i$.
By similarly evaluating
${\bm a}_j^T({\bm{\mathcal E}}_j+{\bm{\mathcal E}}_i){\bm a}_i$,
one finds that
\begin{eqnarray}
{\bm a}_j^T{\bm e}{\bm a}_i &=& \frac{\tilde{E}_i+\tilde{E}_j}{2}
{\bm a}_j^T{\bm a}_i \\ \nonumber
&& +\sum_c \tilde{\gamma}_{ic}\tilde{\gamma}_{jc}
\left(\frac{S_{ic}+S_{jc}}{2}-B_c\right).
\end{eqnarray}
These results are summarized in the matrices ${\bm M}$ and ${\bm N}$:
\begin{equation}
{\bm a}_j^T{\bm a}_i\equiv M_{ij}=\left\{
\begin{array}{ll} 1 & i=j \\ -\sum_c\tilde{\gamma}_{ic}\tilde{\gamma}_{jc}
\frac{S_{ic}-S_{jc}}{\tilde{E}_i-\tilde{E}_j} & i\neq j
\end{array} \right.
\label{eq:mmatrix}
\end{equation}
and
\begin{eqnarray}
\label{eq:nmatrix}
{\bm a}_j^T{\bm e}{\bm a}_i & \equiv & N_{ij} \\ \nonumber
&=& \left\{
\begin{array}{ll}\tilde{E}_i+\sum_c \tilde{\gamma}_{ic}^2(S_{ic}-B_c) & i=j \\
\sum_c \tilde{\gamma}_{ic}\tilde{\gamma}_{jc}
\left(\frac{\tilde{E}_iS_{jc}-\tilde{E}_jS_{ic}}{\tilde{E}_i-\tilde{E}_j}
-B_c \right) & i\neq j \end{array} \right. .
\end{eqnarray}
Note that the construction of ${\bm N}$ requires the adoption
of specific $B_c$ values.

The eigenvectors of Eq.~(\ref{eq:eigen}) can be arranged into
a square matrix ${\bm a}$ such that Eq.~(\ref{eq:newgamma}) becomes
\begin{equation}
\tilde{\bm \gamma}_c= {\bm a}^T {\bm \gamma}_c.
\label{eq:mnewgamma}
\end{equation}
The matrices ${\bm M}$ and ${\bm N}$ defined above can then be written as
${\bm M}={\bm a}^T{\bm a}$ and  ${\bm N}={\bm a}^T{\bm e}{\bm a}$.
From Eq.~(\ref{eq:little_e}) the matrix ${\bm e}$ trivially
satisfies the eigenvalue equation
\begin{equation}
{\bm e} {\bm u}_\lambda = E_\lambda {\bm u}_\lambda.
\end{equation}
Upon substitution of ${\bm u}_\lambda={\bm a} {\bm b}_\lambda$ and
multiplying from the left by ${\bm a}^T$ this equation becomes
\begin{equation}
{\bm N} {\bm b}_\lambda = E_\lambda {\bm M} {\bm b}_\lambda.
\label{eq:master}
\end{equation}
This eigenvalue equation holds the key for transforming from the
$\tilde{E}_i$-$\tilde{\gamma}_{ic}$ representation to the
standard ${\bm R}$-matrix parameters $E_\lambda$ and $\gamma_{\lambda c}$.
The real, symmetric, and energy-independent matrices
${\bm M}$ and ${\bm N}$ are completely
determined by $\tilde{E}_i$, $\tilde{\gamma}_{ic}$, and $B_c$ using
Eqs.~(\ref{eq:mmatrix},\ref{eq:nmatrix}).
The $E_\lambda$ can thus be determined by finding the eigenvalues
of a generalized eigenvalue equation.
If the matrix ${\bm M}$ is also positive definite then Eq.~(\ref{eq:master})
is known as the symmetric-definite eigenvalue problem
and has $N_\lambda$ real eigenvalues (see Sec.~8.7 of Ref.~\cite{Gol96}).
The off-diagonal elements of ${\bm M}$ are
$\approx -\sum_c \tilde{\gamma}_{ic}\tilde{\gamma}_{jc}\frac{dS_c}{dE}$
which is typically small compared to unity; ${\bm M}$ will be
positive definite provided the $\tilde{\gamma}_{ic}$
are not too large and the energy dependences of $S_c$ are not too great.
Further if ${\bm M}$ is not positive definite, the eigenvectors
${\bm a}_i$ are not real and the transformation to standard
${\bm R}$-matrix parameters is not defined.
We thus conclude that for physically reasonable $\tilde{\gamma}_{ic}$,
$\tilde{E}_i$, and $S_{ic}$  the matrix ${\bm M}$ will be positive definite;
in practice we have found this condition to be easily fulfilled.
Finally note that ${\bm M}$ is automatically positive definite for any
given set of standard parameters since ${\bm M}={\bm a}^T{\bm a}$.

The eigenvectors of Eq.~(\ref{eq:master}) ${\bm b}_\lambda$ can be
arranged into a square matrix ${\bm b}$ which satisfies the relations
\begin{equation}
{\bm b}^T {\bm M} {\bm b} = {\bm 1}
\end{equation}
and
\begin{equation}
{\bm b}^T {\bm N} {\bm b} = {\bm e}.
\end{equation}
We therefore have ${\bm b}={\bm a}^{-1}$ and from Eq.~(\ref{eq:mnewgamma})
the standard ${\bm R}$-matrix reduced widths are given by
\begin{equation}
{\bm \gamma}_c = {\bm b}^T \tilde{\bm \gamma}_c .
\label{eq:conv_gamma}
\end{equation}
The simultaneous diagonalization of ${\bm M}$ and ${\bm N}$
thus provides all of the standard ${\bm R}$-matrix parameters.
Note that any $B_c$ can be chosen; the collision matrix ${\bm U}$
will be invariant provided the same $B_c$ are used in
Eqs.~(\ref{eq:nmatrix}) and~(\ref{eq:u_channel}) or~(\ref{eq:ainv}).
The numerical solution of Eq.~(\ref{eq:master}) is discussed in
Sec.~8.7.2 of Ref.~\cite{Gol96}; we have
have utilized the LAPACK~\cite{lapack} routine {\sc dsygv}.

\section{Further Development}
\label{sec:further}

It is fruitful to investigate alternative forms for the level matrix
and the ${\bm R}$ matrix which allow the collision matrix to be calculated
directly from the alternative parameters.

\subsection{The Alternative Level Matrix}
\label{subsec:altlev}

We define the alternative level matrix $\tilde{\bm A}$ implicitly via
\begin{equation}
{\bm \gamma}_{c'}^T {\bm A} {\bm \gamma}_c \equiv
\tilde{\bm \gamma}_{c'}^T \tilde{\bm A} \tilde{\bm \gamma}_c.
\label{eq:def_alta}
\end{equation}
In order for this relation to hold, we must have
\begin{equation}
{\bm a} \tilde{\bm A} {\bm a}^T = {\bm A},
\end{equation}
or equivalently
\begin{equation}
\tilde{\bm A}^{-1}={\bm a}^T {\bm A}^{-1} {\bm a}
\end{equation}
where we have used Eq.~(\ref{eq:mnewgamma}).
We can now substitute Eq.~(\ref{eq:ainv}) for ${\bm A}^{-1}$ and
again use Eq.~(\ref{eq:mnewgamma}) to obtain
\begin{eqnarray}
\tilde{\bm A}^{-1}&=&{\bm a}^T{\bm e}{\bm a} -E{\bm a}^T{\bm a} \\ \nonumber
&&  -\sum_c \tilde{\bm \gamma}_c \tilde{\bm \gamma}_c^T(S_c-B_c+{\rm i}P_c) \\
&=& {\bm N} - E{\bm M}
  -\sum_c \tilde{\bm \gamma}_c \tilde{\bm \gamma}_c^T(S_c-B_c+{\rm i}P_c) .
\end{eqnarray}
The elements of this matrix can now be determined entirely from
the alternative parameters with
the aid of Eqs.~(\ref{eq:mmatrix},\ref{eq:nmatrix}):
\begin{eqnarray}
\label{eq:alta_inv}
(\tilde{\bm A}^{-1})_{ij} &=& (\tilde{E}_i-E)\delta_{ij}-\sum_c
\tilde{\gamma}_{ic} \tilde{\gamma}_{jc} (S_c+{\rm i}P_c) \\ \nonumber
&+& \sum_c \left\{ \begin{array}{ll}
\tilde{\gamma}_{ic}^2 S_{ic} & i=j \\
\tilde{\gamma}_{ic} \tilde{\gamma}_{jc}
\frac{S_{ic}(E-\tilde{E}_j) - S_{jc}(E-\tilde{E}_i)}
{\tilde{E}_i-\tilde{E}_j} & i\neq j \end{array} \right. .
\end{eqnarray}
Note that the boundary-condition constants $B_c$ have now canceled out -- a
not unexpected result since the alternative parameters
and the collision matrix are independent of $B_c$.
We can now express the collision matrix directly in terms of
the alternative parameters using Eqs.~(\ref{eq:u_a},\ref{eq:def_alta})
\begin{equation}
U_{c'c}=\Omega_{c'}\Omega_c\left[ \delta_{c'c}+2{\rm i}(P_{c'} P_c)^{1/2}
\tilde{\bm \gamma}_{c'}^T \tilde{\bm A} \tilde{\bm \gamma}_c \right] .
\label{eq:u_alta}
\end{equation}

\subsection{The Alternative ${\bm R}$ Matrix}
\label{subsec:altr}

The matrix $\tilde{\bm R}$ is an alternative to
the standard ${\bm R}$ matrix and is defined implicitly via
\begin{equation}
[{\bm 1}-{\bm R}({\bm L}-{\bm B})]^{-1}{\bm R} \equiv
({\bm 1}-{\rm i}\tilde{\bm R}{\bm P})^{-1}\tilde{\bm R},
\label{eq:def_altr}
\end{equation}
where ${\bm P}$ is a purely diagonal matrix with elements $P_c$.
By comparison with Eqs.~(\ref{eq:u_channel},\ref{eq:u_alta})
we must have
\begin{equation}
({\bm 1}-{\rm i}\tilde{\bm R}{\bm P})^{-1}\tilde{\bm R} =
\tilde{\bm \gamma}^T \tilde{\bm A} \tilde{\bm \gamma}.
\label{eq:alt_ar}
\end{equation}
We proceed by assuming that $\tilde{\bm R}$ can be written in the form
\begin{equation}
\tilde{\bm R} = \tilde{\bm \gamma}^T {\bm Q} \tilde{\bm \gamma}.
\label{eq:defq}
\end{equation}
In the Appendix we describe a method to derive
the level matrix form for the collision matrix [Eq.~(\ref{eq:u_a_mat})]
from the channel matrix form [Eq.~(\ref{eq:u_channel})].
This reasoning can also be applied to $\tilde{\bm R}$ and $\tilde{\bm A}$.
We find that in order to satisfy Eq.~(\ref{eq:alt_ar}) we must have
\begin{equation}
{\bm Q}^{-1} = \tilde{\bm A}^{-1}+
{\rm i}\tilde{\bm \gamma}{\bm P}\tilde{\bm \gamma}^T.
\label{eq:qinv}
\end{equation}
A formula for the elements of ${\bm Q}^{-1}$ in terms of the
alternative parameters can then be found using
Eqs.~(\ref{eq:alta_inv},\ref{eq:qinv}):
\begin{eqnarray}
\label{eq:q_inv}
(\tilde{\bm Q}^{-1})_{ij} &=& (\tilde{E}_i-E)\delta_{ij}-\sum_c
\tilde{\gamma}_{ic} \tilde{\gamma}_{jc} S_c \\ \nonumber
&+& \sum_c \left\{ \begin{array}{ll}
\tilde{\gamma}_{ic}^2 S_{ic} & i=j \\
\tilde{\gamma}_{ic} \tilde{\gamma}_{jc}
\frac{S_{ic}(E-\tilde{E}_j) - S_{jc}(E-\tilde{E}_i)}
{\tilde{E}_i-\tilde{E}_j} & i\neq j \end{array} \right. .
\end{eqnarray}
Using Eqs.~(\ref{eq:u_channel},\ref{eq:def_altr}) the collision
matrix can now be written as
\begin{equation}
{\bm U}=2{\rm i}{\bm \rho}^{1/2}{\bm O}^{-1}({\bm 1}-{\rm i}\tilde{\bm R}
{\bm P})^{-1} \tilde{\bm R}{\bm \rho}^{1/2}{\bm O}^{-1} +
{\bm I}{\bm O}^{-1} .
\label{eq:u_altr}
\end{equation}
With the $\tilde{\bm R}$ matrix defined by Eqs.~(\ref{eq:defq},\ref{eq:q_inv})
this equation also gives ${\bm U}$ in terms
of the alternative parameters without reference to the
boundary condition constants.

\subsection{Relative merits of $\tilde{\bm R}$ and $\tilde{\bm A}$}

The $\tilde{\bm R}$ matrix is more complicated than ${\bm R}$ and the
calculation of ${\bm U}$ via Eq.~(\ref{eq:u_altr}) requires inverting a
real $N_\lambda\times N_\lambda$ matrix in addition to a
complex $N_c \times N_c$ matrix.
When calculating ${\bm U}$ via the alternative level matrix one
must invert a single complex $N_\lambda\times N_\lambda$ matrix~--
using the alternative ${\bm R}$ matrix approach may thus offer a modest
computational advantage in comparison when $N_\lambda \gg N_c$.
Note however that if it is necessary to calculate ${\bm U}$
for several energies and $N_\lambda >N_c$
it will probably be more computationally efficient to diagonalize
Eq.~(\ref{eq:master}) once and then use the standard ${\bm R}$-matrix
parameters in Eq.~(\ref{eq:u_channel}) to calculate ${\bm U}$,
as Eq.~(\ref{eq:u_channel}) only requires inverting a single
complex $N_c\times N_c$ matrix.

We would also like to point out that this alternative parameterization,
using $\tilde{E}_i$ and $\tilde{\gamma}_{ic}$ with Eq.~(\ref{eq:u_alta})
or~(\ref{eq:u_altr}), may be of formal interest since no
arbitrary boundary condition constants are required, but the equations are
mathematically equivalent to the standard ${\bm R}$-matrix approach.
The alternative parameters in fact correspond to eigenfunctions
satisfying energy-dependent boundary conditions --
the real part of the Kapur-Peierls or Siegert boundary
conditions see (LT, Sec.~IX.2).

\section{Solution of the Nonlinear Eigenvalue Equation}

\begin{table}[tb]
\caption{Standard ${\bm R}$-matrix parameters from Table~III
of Ref.~\protect\cite{Azu94} which describe $J^\pi=1^-$ reactions
in the ${}^{16}{\rm O}$ system, and the alternative parameters
derived from them.
The channel labels $\alpha$ and $\gamma$ describe
${}^{12}{\rm C}+\alpha$ and ${}^{16}{\rm O}+\gamma$, respectively.
The channel radius is 6.5~fm and the boundary condition constant is
chosen so that the level shift vanishes for $E=E_1$.
The $\beta$-decay feeding amplitudes
${\mathcal B}_\lambda$ are equivalent to the quantities
$A_{\lambda 1} \gamma_{\lambda 1}^{-1} N_\alpha^{-1/2}$ of
Ref.~\cite{Azu94}.}
\begin{ruledtabular}
\begin{tabular}{lddd}
$\lambda$ & \multicolumn{1}{c}{1} & \multicolumn{1}{c}{2} &
  \multicolumn{1}{c}{3} \\ \hline
$E_\lambda$ (MeV) & -0.0451 & 2.845 & 11.71 \\
$\gamma_{\lambda\alpha}$ (MeV$^{1/2}$) & 0.0793 & 0.330 & 1.017 \\
$\gamma_{\lambda\gamma}$ (MeV$^{-1}$) &
  \multicolumn{1}{c}{$8.76\times 10^{-6}$} &
  \multicolumn{1}{c}{$-2.44\times 10^{-6}$} &
  \multicolumn{1}{c}{$-2.82\times 10^{-6}$} \\
${\mathcal B}_\lambda$ & 1.194 &  0.558 & -0.629 \\ \hline
$\tilde{E}_\lambda$ (MeV) & -0.0451 & 2.400 & 8.00 \\
$\tilde{\gamma}_{\lambda\alpha}$ (MeV$^{1/2}$) & 0.0793 & 0.471 & 0.912 \\
$\tilde{\gamma}_{\lambda\gamma}$ (MeV$^{-1}$) &
  \multicolumn{1}{c}{$8.76\times 10^{-6}$} &
  \multicolumn{1}{c}{$-3.20 \times 10^{-6}$} &
  \multicolumn{1}{c}{$-2.50\times 10^{-6}$} \\
$\tilde{\mathcal B}_\lambda$ & 1.194 & 0.408 & -0.781
\end{tabular}
\end{ruledtabular}
\label{tab:params}
\end{table}

The transformation from $\tilde{E}_{ic}$ and $\tilde{\gamma}_{ic}$ to
the standard ${\bm R}$-matrix parameters $E_\lambda$ and $\gamma_{\lambda c}$
can be carried out in a straightforward manner using the
methods discussed above in Subsec.~\ref{subsec:rel_stand}.
We will now discuss the inverse transformation, i.e. the solution
of the nonlinear eigenvalue problem Eq.~(\ref{eq:eigen}).
At this point it is instructive to introduce a concrete example:
in Table~\ref{tab:params} we show a simple well-documented set of standard
${\bm R}$-matrix parameters taken from Azuma {\em et al.}~\cite{Azu94}.

We consider the {\em linear} eigenvalue equation
\begin{equation}
{\bm {\mathcal E}}(E)\hat{\bm a}_i = \hat{E}_i\hat{\bm a}_i
\label{eq:peigen}
\end{equation}
where ${\bm {\mathcal E}}$, the eigenvalues $\hat{E}_i$, and eigenvectors
$\hat{\bm a}_i$ depend upon on the energy parameter $E$. 
The solutions to the original nonlinear problem Eq.~(\ref{eq:eigen})
thus occur when $\hat{E}_i(E)=E$ in which case $E=\tilde{E}_i$.
From inspection of Eqs.~(\ref{eq:cmatrix},\ref{eq:ematrix})
we see that the eigenvalues $\hat{E}_i$ also correspond to a set of standard
${\bm R}$-matrix level energies, transformed from the original
parameter values to $B_c=S_c(E)$.

We will next investigate how the $\hat{E}_i$ depend on $E$.
Starting with
\begin{equation}
\hat{E}_i=\hat{\bm a}_i^T{\bm{\mathcal E}}\hat{\bm a}_i
\label{eq:eigen_param}
\end{equation}
differentiation with respect to $E$ yields
\begin{eqnarray}
\frac{d\hat{E}_i}{dE}&=&
  \hat{\bm a}_i^T\frac{d{\bm{\mathcal E}}}{dE}\hat{\bm a}_i +
  \frac{d\hat{\bm a}_i^T}{dE}{\bm{\mathcal E}}\hat{\bm a}_i +
  \hat{\bm a}_i^T{\bm{\mathcal E}}\frac{d\hat{\bm a}_i}{dE} \\
&=& -\hat{\bm a}_i^T\left(\sum_c {\bm \gamma}_c{\bm \gamma_c}^T \,
  \frac{dS_c}{dE}\right)\hat{\bm a}_i \\ \nonumber
&&   +\hat{E}_i \left(  \frac{d\hat{\bm a}_i^T}{dE}
 \hat{\bm a}_i + \hat{\bm a}_i^T \, \frac{d\hat{\bm a}_i}{dE} \right).
\end{eqnarray}
Since by definition $\hat{\bm a}_i^T\hat{\bm a}_i=1$ we have
$\frac{d\hat{\bm a}_i^T}{dE}
  \hat{\bm a}_i + \hat{\bm a}_i^T \, \frac{d\hat{\bm a}_i}{dE}=0$
and we finally find
\begin{equation}
\frac{d\hat{E}_i}{dE}=-\sum_c ({\bm \gamma}_c^T \hat{\bm a}_i)^2
  \frac{dS_c}{dE}.
\label{eq:deig_de}
\end{equation}

The energy derivative of the shift function $\frac{dS_c}{dE}$
is positive for negative-energy channels, and is $\ge 0$
for positive-energy channels for all cases we are aware of.
This point is also discussed by (LT, p.~350);
although a general proof of $\frac{dS_c}{dE}\ge 0$ is lacking
it appears to always hold in practice and we will thus assume
it is true here. Note that for any specific case it is a simple matter
to verify this relation numerically.

\begin{figure}[tb]
\begin{center}
\includegraphics[bbllx=90,bblly=23,bburx=555,bbury=722,%
angle=-90,width=3.4in]{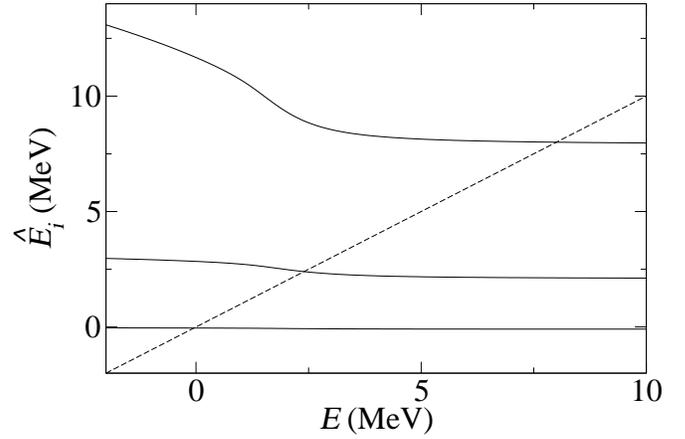}
\end{center}
\caption{The eigenvalue trajectories are shown by plotting as solid curves
the eigenvalues $\hat{E}_i$ versus $E$;
the dashed line corresponds to $\hat{E}_i=E$.
The eigenvalues of Eq.~(\protect\ref{eq:eigen}) are given by the
intersections between the solid curves and the dashed line.}
\label{fig:eigen_traj}
\end{figure}

Since $({\bm \gamma}_c^T \hat{\bm a}_i)^2$ is clearly $\ge 0$,
we can utilize $\frac{dS_c}{dE}\ge 0$ to conclude
from Eq.~(\ref{eq:deig_de}) that
\begin{equation}
\frac{d\hat{E}_i}{dE} \le 0.
\label{eq:eig_slope}
\end{equation}
The eigenvalues $\hat{E}_i$ are thus
monotonically non-increasing functions of $E$.
The eigenvalue trajectories for the example parameters are shown
in Fig.~\ref{fig:eigen_traj} where the expected behavior is seen.
We also note the the eigenvalue trajectories avoid crossing one another for
the reasons given by von Neumann and Wigner~\cite{Neu29}. 
The avoided-crossing behavior is most apparent when there are two
nearby levels with very different reduced width amplitudes.

The nonlinear eigenvalue problem Eq.~(\ref{eq:eigen}) and
the parametric eigenvalue problem Eq.~(\ref{eq:eigen_param}) are
also closely related to a well-studied question in linear algebra:
the modification of a symmetric matrix with known eigenvalues and
eigenvectors by a positive-definite perturbation.
This question is analyzed for the single-channel case in
Sec.~8.5.3 of Golub and Van~Loan~\cite{Gol96} and for
the multi-channel case  by Arbenz, Gander, and Golub~\cite{Arb88}.
The perturbation bounds on the eigenvalues derived in
Ref.~\cite{Arb88} imply that the $\hat{E}_i$ remain finite
provided the $S_c$ are finite -- thus the eigenvalue trajectories
do not have real poles for $|E|<\infty$.

From the lack of poles and the monotonic dependence
$\frac{d\hat{E}_i}{dE} \le 0$ we can  conclude that each
eigenvalue trajectory intersects with the line $\hat{E}_i=E$ exactly once.
These intersections are shown graphically
for the example in Fig.~\ref{fig:eigen_traj}.
We thus have the important result that the
non-linear eigenvalue problem Eq.~(\ref{eq:eigen}) has a number of
real eigenvalues exactly equal to the number of ${\bm R}$-matrix levels.
A similar type of nonlinear eigenvalue problem has been investigated by
Rogers~\cite{Rog64}; it may be that the methods described in that paper
could be used to develop further understanding of the present problem,
e.g. to investigate inner products and/or the linear independence of
the eigenvectors.

The eigenvalues of Eq.~(\ref{eq:eigen}) satisfy the characteristic equation
\begin{equation}
{\rm det}({\bm {\mathcal E}}-E{\bm 1})=0
\end{equation}
which can also be written as
\begin{equation}
{\rm det}\left[{\bm e}-E{\bm 1}-{\bm \gamma}
  ({\bm S}-{\bm B}){\bm \gamma}^T\right]=0,
\label{eq:eq_solv1}
\end{equation}
where ${\bm S}$ is a purely diagonal matrix with elements $S_c$
which depend upon $E$.
Using the methods described in Ref.~\cite{Arb88} one can show that
\begin{eqnarray}
\lefteqn{ {\rm det}\left[{\bm e}-E{\bm 1}-{\bm \gamma}
  ({\bm S}-{\bm B}){\bm \gamma}^T \right]=
  {\rm det}({\bm e}-E{\bm 1}) }\hspace{0.15in} \\ \nonumber &&
  \times\,\, {\rm det}\left[ {\bm 1}-{\bm \gamma}^T
  ({\bm e}-E{\bm 1})^{-1}{\bm \gamma}
  ({\bm S}-{\bm B}) \right] .
\end{eqnarray}
The eigenvalues thus satisfy
\begin{equation}
 {\rm det}({\bm e}-E{\bm 1})\,{\rm det}\left[ {\bm 1}-{\bm R}
  ({\bm S}-{\bm B}) \right]=0,
\label{eq:eq_solv2}
\end{equation}
which may be a computationally-efficient approach for determining the
eigenvalues when $N_\lambda>N_c$
since the calculation of ${\rm det}({\bm e}-E{\bm 1})$ is trivial.
Note that Eq.~(\ref{eq:eq_solv2}) is the multi-channel arbitrary-$B_c$
generalization of the resonance condition given by
Eq.~(14) of Ref.~\cite{Ang00}.
The eigenvalues also satisfy
\begin{equation}
 {\rm det}\left[ {\bm 1}-{\bm R}({\bm S}-{\bm B}) \right]=0,
\end{equation}
but this equation has poles in addition to zeros, and
if there is a level $\lambda$ with $B_c=S_c(E_\lambda)$
(at most one level can satisfy this condition) it does not produced a zero.

\begin{table}[tb]
\caption{Elements $b_{ij}$ of the transformation matrix ${\bm b}$
corresponding to the parameters of Table~\ref{tab:params}.}
\begin{ruledtabular}
\begin{tabular}{cddd}
\backslashbox{$i$}{$j$}  & \multicolumn{1}{c}{1} & \multicolumn{1}{c}{2} &
  \multicolumn{1}{c}{3}      \\ \hline
1  & 1.000 & 0.0373  & 0.0446 \\
2  & 0.000 & 0.9781  & 0.2281 \\
3  & 0.000 & -0.1466 & 0.9933
\end{tabular}
\end{ruledtabular}
\label{tab:b_elements}
\end{table}

Rather than finding the eigenvalues by directly solving the characteristic
equation, we have applied the Rayleigh Quotient Iteration method described
in Sec.~8.2.3 of Ref.~\cite{Gol96} to Eq.~(\ref{eq:eigen}),
as this procedure yields the eigenvectors as well as eigenvalues.
Starting values for the eigenvalues and eigenvectors were taken as
$\tilde{E}_i=E_i+\sum_c\gamma_{i c}^2[S_c(E_i)-B_c]$ and $a_{ji}=\delta_{ji}$.
Due to the nonlinear nature of the problem, the matrix
${\bm {\mathcal E}}$ must be updated at each step of the iteration.
These procedures were tested with
several single-channel and multi-channel parameter sets, and
were successful in finding all of the eigenvalues in every case.
We cannot rule out however that some cases may require
more carefully chosen starting values.
Once the $\tilde{E}_i$ and ${\bm a}_i$ are found,
the $\tilde{\gamma}_{ic}$ can be calculated using Eq.~(\ref{eq:newgamma}).
In Table~\ref{tab:params} we show for the example case the alternative
parameters determined from the standard ${\bm R}$-matrix parameters.
Note that the alternative parameters are exactly
the same as the ${\bm R}$-matrix parameters given in the last column of
Table~III of Ref.~\cite{Azu94} which have been transformed
to satisfy $B_c=S_c(E_\lambda)$ for other levels.
As discussed in Subsec.~\ref{subsec:def_alt} this equality is
required due to our definition of the alternative parameters.
In Table~\ref{tab:b_elements} we show the elements of the
matrix ${\bm b}$ for the example parameters.
Finally we would like to point out that the methods discussed in
this section should be generally useful for the extraction of resonance
parameters from standard ${\bm R}$-matrix parameters.

\section{Application to ${\bm \gamma}$~Rays and ${\bm \beta}$~Decays}

We will briefly discuss the application of the alternative
parameterization to reactions involving $\gamma$ rays and $\beta$ decays.
Gamma-ray decay processes are generally treated with first-order
perturbation theory in ${\bm R}$-matrix theory,
which implies that $\gamma$-ray channels are excluded from the sum over
channels when constructing ${\bm A}$, $\tilde{\bm A}$,
${\bm M}$, or ${\bm N}$.
Assuming that external contributions can be ignored, the collision
matrix elements connecting $\gamma$-ray channels (labeled $\gamma$)
and non-$\gamma$-ray channels (labeled $c$) are
given by (LT, Eq.~XIII.3.9)
\begin{eqnarray}
U_{c\gamma} &=& 2{\rm i}\Omega_c (P_cP_\gamma)^{1/2}
\sum_{\lambda\mu} \gamma_{\lambda c}\gamma_{\mu \gamma}A_{\lambda\mu}\\
\nonumber
&=& 2{\rm i}\Omega_c (P_cP_\gamma)^{1/2}
{\bm \gamma}_c^T {\bm A} {\bm \gamma}_\gamma .
\end{eqnarray}
In the long-wavelength approximation the penetration factor
for $\gamma$ rays is given by $P_\gamma =E_\gamma^{2\ell+1}$
where $\ell$ is the multipolarity.
The observed $\gamma$-ray widths are described by Eq.~(\ref{eq:Gamma}),
where $\gamma$-ray channels are excluded from the sum in the denominator.
Using the same reasoning described in Subsec.~\ref{subsec:altlev}
the alternative expression for the collision matrix elements can
be obtained using the replacement
\begin{equation}
{\bm \gamma}_c^T {\bm A}{\bm \gamma}_\gamma =
\tilde{\bm \gamma}_c^T \tilde{\bm A}\tilde{\bm \gamma}_\gamma ,
\end{equation}
where the alternative $\gamma$-ray reduced width amplitudes are related
to the standard parameters via
\begin{equation}
{\bm \gamma}_\gamma = {\bm b}^T \tilde{\bm \gamma}_\gamma .
\end{equation}
If the external contributions to the matrix elements are
included using the formalism of Barker and Kajino~\cite{Bar91},
the expressions for the collision matrix elements and observed widths
become more complicated. However these quantities can still be
written in terms of the alternative parameters using the above
equations, noting that the $\gamma_{\lambda\gamma}$ above are
the {\em internal} $\gamma$-ray reduced width amplitudes.

The extension of the alternative parameterization to the description
of $\beta$-delayed particle spectra is straightforward.
A multi-channel formula for the particle spectrum
is given by Eq.~(7) of Barker and Warburton~\cite{Bar88};
note that additional parameters must now be introduced,
the $\beta$-decay feeding amplitudes ${\mathcal B}_{\lambda x}$.
It is convenient to form column vectors
${\bm {\mathcal B}}_x$ from the ${\mathcal B}_{\lambda x}$, so
that $\sum_{\lambda\mu} {\mathcal B}_{\lambda x} A_{\lambda\mu}\gamma_{\mu c}$
can be written as ${\bm \gamma}_c^T {\bm A}{\bm {\mathcal B}}_x$.
Again using the reasoning of Subsec.~\ref{subsec:altlev} we have
\begin{equation}
{\bm \gamma}_c^T {\bm A}{\bm {\mathcal B}}_x =
\tilde{\bm \gamma}_c^T \tilde{\bm A}\tilde{\bm {\mathcal B}}_x,
\label{eq:alt_feed}
\end{equation}
where the alternative feeding amplitudes $\tilde{\bm {\mathcal B}}_x$
are related to the standard parameters via
\begin{equation}
{\bm {\mathcal B}}_x ={\bm b}^T\tilde{\bm {\mathcal B}}_x.
\label{eq:feed_conv}
\end{equation}
Note also that if $B_c=S_c(E_\lambda)$ we have
$\tilde{\mathcal B}_{\lambda x} = {\mathcal B}_{\lambda x}$.
The $\beta$-delayed particle spectrum can now be calculated directly from
the alternative parameters by using Eq.~(\ref{eq:alt_feed})
in Eq.~(7) of Ref.~\cite{Bar88}. One could also convert to standard
${\bm R}$-matrix parameters using Eq.~(\ref{eq:feed_conv}) and
the methods discussed in Subsec.~\ref{subsec:rel_stand}, 
and then calculate the spectrum using standard ${\bm R}$-matrix formulas.

In Table~\ref{tab:params} we also show the standard and alternative
$\gamma$-ray reduced amplitudes and $\beta$-decay feeding amplitudes
for the example case.

\section{Conclusions}

We have presented an alternative formulation of ${\bm R}$-matrix theory
based on the parameters $\tilde{E}_i$ and $\tilde{\gamma}_{ic}$
defined in Subsec.~\ref{subsec:def_alt}.
This parameterization is a generalization of the ideas presented
by Angulo and Descouvemont~\cite{Ang00}.
The new formulation is mathematically equivalent to the standard
${\bm R}$-matrix theory~\cite{Lan58} but there are no boundary condition
constants or level shifts. The new parameters can be converted to
standard ${\bm R}$-matrix parameters by diagonalizing Eq.~(\ref{eq:master}),
or be used to calculate the collision matrix directly using
Eqs.~(\ref{eq:u_alta}) or~(\ref{eq:u_altr}).
We have discussed the solution of the nonlinear eigenvalue problem
Eq.~(\ref{eq:eigen}) which is needed to convert standard ${\bm R}$-matrix
parameters to the new parameterization.
Finally we have briefly discussed the application to $\gamma$~rays
and $\beta$ decays.

We can envision at least two uses for this new formulation in the
fitting of experimental data. One application is the generation
of starting parameter values from an outside source of spectroscopic
information such as a level compilation or shell-model calculation.
These latter sources generally do not supply standard ${\bm R}$-matrix
parameters but rather resonance parameters without level shifts.
In the past the methods to incorporate these types
of information have not always been optimal
(e.g. $B_c$ could be chosen to make the level shift vanish
for a representative energy, but not for all energies simultaneously).
Another application is to use the alternative parameters as the fit parameters.
The calculations can be made directly from the alternative parameters
using the methods discussed in Sec.~\ref{sec:further},
or by diagonalizing Eq.~(\ref{eq:master})
to find the standard ${\bm R}$-matrix parameters.
The latter option may be preferable if $N_\lambda > N_c$,
if observables must be calculated for many different energies.
It should be noted that in data-fitting applications the collision matrix
must be calculated repeatedly for different energies, and
the extra computational overhead required will be negligible in comparison.
With the alternative parameters it is very easy to
fix known information about level energies and partial widths
for any number of levels.

\begin{acknowledgments}

It is a pleasure to thank Fred Barker, Pierre Descouvemont, and Gerry Hale,
for useful discussions.
We also thank the Institute for Nuclear Theory (INT) at the University of
Washington for its hospitality during a part of this work.
Financial support was supplied in part by the U.S. Department of Energy,
through the INT and also Grant No. DE-FG02-88ER40387.

\end{acknowledgments}

\appendix*
\section{}
\label{sec:app}

The equivalence of the two forms of the collision matrix
given by Eqs.~(\ref{eq:u_channel},\ref{eq:u_a_mat}) is discussed
in (LT, Sec.~IX.1). The derivation is reviewed here, utilizing
the matrix notation introduced in Sec.~\ref{sec:standardR}.
The same procedure is useful for the derivation of
the alternative ${\bm R}$ matrix as discussed in Subsec.~\ref{subsec:altr}.

We define ${\bm L}_0={\bm L}-{\bm B}$ and note
the quantity $[{\bm 1}-{\bm R}({\bm L}-{\bm B})]^{-1}$
in Eq.~(\ref{eq:u_channel}) can be written as
\begin{eqnarray}
\lefteqn{ ({\bm 1}-{\bm R}{\bm L}_0)^{-1}=} \hspace{0.25in} && \\ \nonumber
&& [{\bm L}_0-({\bm L}_0 {\bm \gamma}^T)({\bm e}-E{\bm 1})^{-1}
({\bm \gamma}{\bm L}_0) ]^{-1}{\bm L}_0.
\end{eqnarray}
A useful identity is given by
\begin{eqnarray}
\lefteqn{ ({\bm X}+{\bm Z}{\bm Y}{\bm Z}^T)^{-1} =
  {\bm X}^{-1} }\hspace{0.25in} && \\ \nonumber
  && -{\bm X}^{-1}{\bm Z}({\bm Y}^{-1}+
  {\bm Z}^T {\bm X}^{-1} {\bm Z})^{-1} {\bm Z}^T {\bm X}^{-1} ,
\end{eqnarray}
which holds for any square and invertible matrices ${\bm X}$ and ${\bm Y}$
which need not be of the same dimension~\cite{Zha99}.
With the aid of this identity we obtain
\begin{eqnarray}
\lefteqn{ ({\bm 1}-{\bm R}{\bm L}_0)^{-1} =
\biggl\{ {\bm L}_0^{-1}-{\bm L}_0^{-1}({\bm L}_0{\bm \gamma}^T)
\times \biggr.} \hspace{0.1in} && \\ \nonumber
&& \left [-({\bm e}-E{\bm 1}) +({\bm \gamma}{\bm L}_0) {\bm L}_0^{-1}
({\bm L}_0{\bm \gamma}^T)\right ]^{-1}
({\bm \gamma}{\bm L}_0){\bm L}_0^{-1} \biggl\}{\bm L}_0 \\ 
&& \hspace{0.25in}
= {\bm 1}+{\bm \gamma}^T({\bm e}-E{\bm 1}-{\bm \gamma}{\bm L}_0
{\bm \gamma}^T)^{-1}{\bm \gamma}{\bm L}_0 \\
&& \hspace{0.25in}
= {\bm 1}+{\bm \gamma}^T{\bm A}{\bm \gamma}{\bm L}_0 ,
\end{eqnarray}
where in the last step we have used Eq.~(\ref{eq:ainv}) for the
definition of the level matrix ${\bm A}$.

We can then write
\begin{eqnarray}
({\bm 1}-{\bm R}{\bm L}_0)^{-1}{\bm R} &=&
({\bm 1}+{\bm \gamma}^T{\bm A}{\bm \gamma}{\bm L}_0)
\times\hspace{0.75in} \\ \nonumber
&& {\bm \gamma}^T({\bm e}-E{\bm 1})^{-1}{\bm \gamma} \\
&=&{\bm \gamma}^T({\bm e}-E{\bm 1})^{-1}{\bm \gamma} \\ \nonumber
\lefteqn{ + {\bm \gamma}^T{\bm A}[ -{\bm A}^{-1} +({\bm e}-E{\bm 1})]
({\bm e}-E{\bm 1})^{-1}{\bm \gamma} } \hspace{0.5in} &&
\end{eqnarray}
where we have substituted $-{\bm A}^{-1} +({\bm e}-E{\bm 1})$ for
${\bm \gamma}{\bm L}_0{\bm \gamma}^T$.
Simplifying this expression we finally have
\begin{equation}
({\bm 1}-{\bm R}{\bm L}_0)^{-1}{\bm R} = {\bm \gamma}^T{\bm A}{\bm \gamma}
\end{equation}
which proves the equivalence of
Eqs.~(\ref{eq:u_channel},\ref{eq:u_a_mat}).

\end{document}